\documentclass[aps,prc,twocolumn,superscriptaddress,nofootinbib,tightenlines,
preprintnumbers,showkeys,floatfix]{revtex4-2}

\usepackage[utf8]{inputenc}
\usepackage[english]{babel}
\usepackage{amssymb,amsthm,amsmath,amstext,amsbsy,amsopn,mathrsfs}
\usepackage{bbm}
\usepackage{nicefrac}
\usepackage{slashed}
\usepackage{hyperref}
\usepackage{leftidx}
\usepackage{environ}
\usepackage{mathtools}
\usepackage{xspace}
\usepackage{array}
\usepackage{xcolor}
\usepackage{isotope}
\usepackage{suffix}
\usepackage{booktabs,colortbl}
\usepackage{nth}
\usepackage{braket}
\usepackage{overpic}
\usepackage{tikz}
\newcommand{\ie}{\textit{i.e.}\xspace}
\newcommand{\eg}{\textit{e.g.}\xspace}
\newcommand{\cf}{\textit{cf.}\xspace}
\newcommand{\apriori}{\textit{a priori}\xspace}
\WithSuffix\newcommand\apriori*{\textit{a-priori}\xspace}

\newcommand{\dd}{\mathrm{d}}

\newcommand{\ii}{\mathrm{i}}
\newcommand{\ee}{\mathrm{e}}

\newcommand{\RR}{\mathbb{R}}

\newcommand{\Rp}{\mathrm{Re}}
\newcommand{\Ip}{\mathrm{Im}}

\newcommand*\rvec[1]%
{\ensuremath{\overset{\smash{\raisebox{-1.5pt}{\tiny$\rightarrow$}}}{#1}}}
\newcommand*\lvec[1]%
{\ensuremath{\overset{\smash{\raisebox{-1.5pt}{\tiny$\leftarrow$}}}{#1}}}

\NewEnviron{subalign}[1][]{%
\begin{subequations}\begin{align}
  \BODY
\end{align}\label{#1}\end{subequations}
}

\NewEnviron{spliteq}{%
\begin{equation}\begin{split}
  \BODY
\end{split}\end{equation}
}

\NewEnviron{mleq}{%
\begin{equation}
  \BODY
\end{equation}
}

\renewcommand{\vec}[1]{\mathbf{#1}}

\newcommand{\RHAECsubfig}[2]{
  \begin{tikzpicture}
    \draw (0, 0) node[anchor=north west, inner sep=0] {#2};
    \draw (1.7, -2.1) node[anchor=north west] {#1};
  \end{tikzpicture}
}

\begin{document}

\title{Eigenvector continuation for emulating and extrapolating two-body resonances}

\author{Nuwan Yapa}
\email{ysyapa@ncsu.edu}
\affiliation{Department of Physics, North Carolina State University,
Raleigh, North Carolina 27695, USA}

\author{K\'evin Fossez}
\email{kfossez@fsu.edu}
\affiliation{ Department of Physics, Florida State University,
Tallahassee, Florida 32306, USA}
\affiliation{Physics Division, Argonne National Laboratory, Lemont,
Illinois 60439, USA}

\author{Sebastian K\"onig}
\email{skoenig@ncsu.edu}
\affiliation{Department of Physics, North Carolina State University,
Raleigh, North Carolina 27695, USA}

\begin{abstract}
The study of open quantum systems (OQSs), \ie, systems interacting with an environment,
impacts our understanding of exotic nuclei in low-energy nuclear physics,
hadrons, cold-atom systems, or even noisy intermediate-scale quantum computers.
Such systems often exhibit resonance states characterized by energy positions and dispersions (or decay widths), the properties of which can be difficult to predict theoretically due to
their coupling to the continuum of scattering states.
Dealing with this phenomenon poses challenges both conceptually and numerically.
For that reason, we investigate how the reduced basis method known as eigenvector continuation (EC),
which has emerged as a powerful tool to emulate bound and scattering states in closed quantum systems, can be used to study resonance properties.
In particular, we present a generalization of EC that we call \textit{conjugate-augmented eigenvector continuation}, which is based on the complex-scaling method and designed to predict Gamow-Siegert states, and thus resonant properties of OQSs, using only bound-state wave functions as input.
\end{abstract}

\maketitle

\section{Introduction}
\label{sec:Introduction}

Resonances are a ubiquitous phenomenon in physics and are found, for example, in materials,
acoustic devices, or even in planetary motion.
Generally, they appear as amplitude enhancements at the so-called natural frequencies
of the system considered.
As early as 1884, Thomson used complex frequencies to describe the ``decay'' of transient
states in specific electric systems~\cite{thomson84_1276}.
In quantum mechanics, \ie, the appropriate physical theory at microscopic scales, natural
frequencies of a system are associated with ``eigenstates.''
However, the inherently time-dependent nature of resonances makes their formal description as
proper eigenstates delicate.
Indeed, while in scattering theory resonances appear as poles of the scattering ($S$) matrix
and are manifest as peaks in the cross section characterized by an energy position
$E_R$ and a dispersion (or decay width) $\Gamma$, it is only in the quasistationary
formalism~\cite{zeldovich69_b3} that resonances can actually be treated as eigenstates; this
was first realized by Gamow~\cite{gamow28_500} and Siegert~\cite{siegert39_132} in the context
of quantum decay.
In this formalism, the momentum $k$ associated with an eigenstate can be complex, leading to
complex eigenenergies $E = E_R - \ii\Gamma/2$.

Mathematically, quantum states can be divided
into three categories depending on their properties as singularities of the resolvent operator
(full Green's function)~\cite{peierls,chiba15_2645}
\begin{equation}
 G(z) = {(z - H)}^{{-}1} \,,
\end{equation}
where $H$ is the Hamiltonian describing the physical system of interest.
Poles of $G(z)$ located on the negative real axis correspond to bound states, and those located
in the \nth{1} and \nth{4} quadrants of the second Riemann sheet correspond to
so-called resonant states (see details in the next section).
The branch cut of $G(z)$ running along the positive real axis is associated with scattering states.
In contradistinction to bound states, wave functions describing resonant and scattering states
are not square-integrable.
Although for that reason such states do not belong in a Hilbert space, it
is possible to construct a so-called rigged Hilbert space~\cite{bohm97_59} in which quantum
mechanics for all types of states listed above can be formulated rigorously.
The study of quantum resonances thus presents profound conceptual questions and is directly
connected to the fundamental problems of quantum decay~\cite{bohm81_704,bohm03_533} and
irreversibility~\cite{bohm97_64,gadella99_525,castagnino99_527,castagnino06_516,bohm08_198},
as well as to the collapse of the wave functions, all of which lead naturally to the open quantum system (OQS) framework~\cite{fonda78_447,sokolov92_1364,Moiseyev:1998aa,okolowicz03_21,%
civitarese04_34} describing quantum systems coupled to a classical or quantum environment.

Despite their broad relevance, resonances in quantum systems are still challenging to describe
theoretically---and in particular to treat computationally---in many common instances.
The few-body dynamics of resonances involving no more than a handful of particles coupled to the
continuum of scattering states can be described exactly, with state-of-the-art calculations,
based on the Faddeev-Yakubovsky formalism extended to the complex-energy plane using the uniform complex-scaling method,
with the record currently standing at five particles~\cite{lazauskas20_2691,myo20_2674,myo21_2647}.
However, difficulties remain in many-body systems composed of ten or more particles that can feature
resonances involving a few or more particles coupled to the continuum.
For such systems, the options are often limited to quasiexact many-body techniques generalized in the quasistationary formalism, which, in principle, can deal with broad resonances, but are often plagued by an intractable increase of the computational cost, due to the discretization of the continuum, or fail to identify physical states in the complex-energy plane;
or they are limited to lattice and quantum Monte Carlo methods that discretize systems in coordinate space but tend to be limited to narrow resonances, \ie, resonances with $\Gamma \ll E_R$, behaving similarly to bound states and exhibiting effective few-body dynamics.

In this work, as a first step towards addressing the need for stable and scalable calculations of
many-body resonances, we explore the possibility of applying a reduced basis method known as eigenvector continuation (EC) to two-body resonances.
In particular, we construct a technique to perform robust bound-state-to-resonance extrapolations in two-body systems.
The versatile EC method was originally introduced in Ref.~\cite{Frame:2017fah}
and quickly found many applications in low-energy nuclear physics~\cite{%
Demol:2019yjt,Konig:2019adq,Ekstrom:2019lss,franzke22_2644,yoshida22_2569,anderson22_2642}.
Its impressive convergence properties were analyzed in Ref.~\cite{Sarkar:2020mad}.
In particular, EC has been used to build emulators~\cite{melendez22_2639,Bonilla:2022rph,Drischler:2022ipa} in
the context of two-body
scattering~\cite{furnstahl20_2455,drischler21_2453,Bai:2021xok,bai22_2641,Garcia:2023slj}, but so
far it has not been applied directly to resonance states.

In this work, we close this gap.
We start by introducing the general formalism in Sec.~\ref{sec:Formalism} to show how $S$-matrix poles can be
extracted using the uniform complex-scaling technique.
In Sec.~\ref{sec:Results} we then present
the implementation and generalization of EC to perform resonance-to-resonance and bound-state-to-resonance
extrapolations.
We demonstrate all developments with concrete numerical examples.
Finally, we summarize our results in Sec.~\ref{sec:Conclusion}.

\section{Formalism}
\label{sec:Formalism}

Because in this work we are interested primarily in a new conceptual development, we
study a quantum system of two particles with masses $m = 2\mu$ and interacting via a spherically symmetric local potential $V$.
We only require the potential to be short ranged or, more specifically, that the interaction between the two particles falls
off quicker than $\mathcal{O}(r^{-3})$~\cite[p.~27]{Taylor:1972pty} with the relative distance $r$ as $r \to \infty$.

\subsection{Basic scattering theory}
\label{sec:Basics}

We start by collecting relevant results from basic scattering theory.
Throughout the discussion, we use natural units with $\hbar = c = 1$.
Any eigenstate of the quantum system considered has to satisfy the Schr\"odinger equation
\begin{equation}
 \left[H_0 + V - E \right] \ket{\psi} = 0 \,,
\label{eq:SEq-op}
\end{equation}
where $H_0$ is the free Hamiltonian, which in momentum space is given by
${p^2}/{(2\mu)}$ in terms of the momentum operator $p$.
By the assumption of spherical symmetry, the equation can be decomposed into
partial waves and each state $\ket{\psi}$ will have definite angular momentum
$l$.
The energy $E$ in Eq.~\eqref{eq:SEq-op} will be negative real for a bound state
(of which there can be at most a finite set), positive real for a scattering
state, and complex with a positive real part and a negative imaginary part for a
decaying resonance.
We come back to these different cases, and in particular to the treatment of resonances as eigenstates, in more detail below.

Introducing the ``wave number'' $k$ by setting $E = k^2 / (2\mu)$, we obtain from
Eq.~\eqref{eq:SEq-op} the radial Schr\"odinger equation
\begin{equation}
 \left[\frac{\mathrm{d}^2}{\mathrm{d}r^2} - \frac{l(l+1)}{r^2} - 2\mu V(r) + k^2
 \right] \psi_{l,k}(r) = 0 \,,
\label{eq:SEq-rad}
\end{equation}
where $\psi_{l,k}(r)$ is the reduced radial wave function that we define here
via
\begin{equation}
 \braket{\vec{r}|\psi} = 2\ii^l \sqrt{\frac{\mu}{\pi k}} Y_l^m(\hat{r})\frac{\psi_{l,k}(r)}{r} \,,
\label{eq:WF-rad}
\end{equation}
with the standard spherical harmonics $Y_l^m(\hat{r})$ and assuming $\ket{\psi}$
has quantum numbers $(l,m)$.

\begin{figure}
    \centering
    \includegraphics[width=0.49\textwidth]{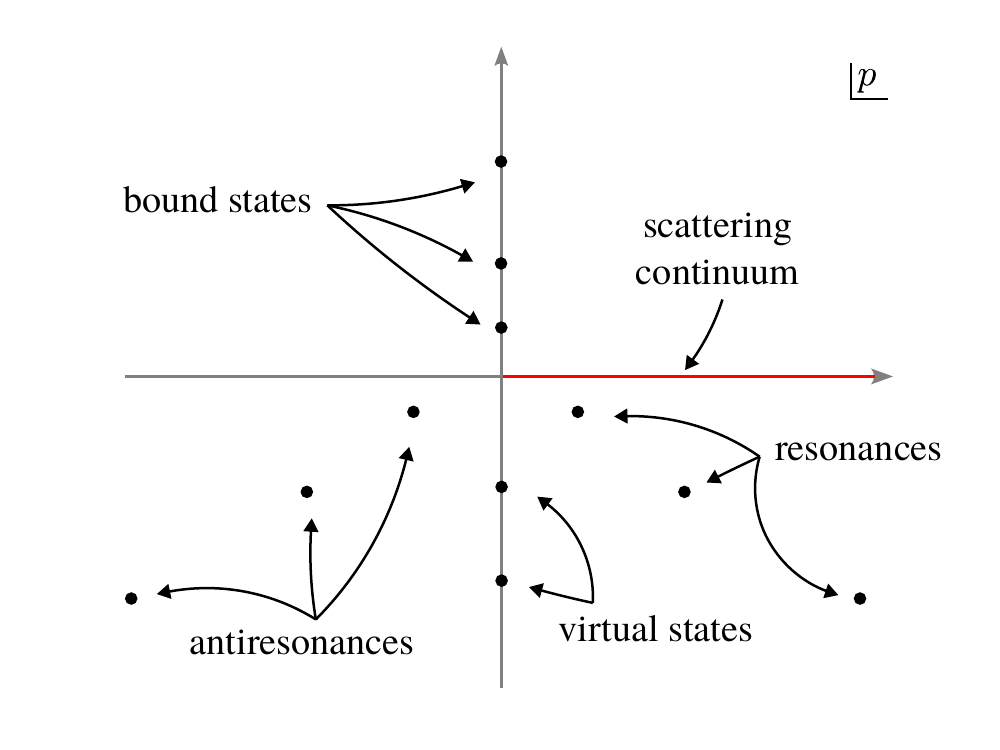}
    \caption{The analytic structure of the $S$ matrix indicating its poles in the $p$ plane corresponding to bound states, virtual states, resonances, and antiresonances (capturing resonances).
    Note especially how bound states lie on the positive imaginary axis while resonances are located in the \nth{4} quadrant.}
    \label{fig:s_matrix}
\end{figure}

By our assumption of a short-range potential, $\psi_{l,k}(r)$ takes the following simple form for asymptotically large $r$:
\begin{equation}
 \label{eq:asymptotic_wf}
 \psi_{l,k}(r) \xrightarrow[r \rightarrow \infty]{}
 \frac{i}{2} \left[\hat{h}^-_l(kr) - s_l(k) \hat{h}^+_l(kr)\right] \,.
\end{equation}
Here $\hat{h}^\pm_l(z)$ are the Riccati-Hankel (RH) functions and $s_l(k)$ is the
partial-wave $S$ matrix, defined implicitly through Eq.~\eqref{eq:asymptotic_wf}.
For more details, we refer to Ref.~\cite{Taylor:1972pty}, from which we have
adopted the conventions used here.
Equation~\eqref{eq:asymptotic_wf} implies that wherever $s_l(k)$ has a pole, we have
\begin{equation}
\label{eq:pobc}
 \psi_{l,k}(r) \xrightarrow[r \rightarrow \infty]{} N \, \hat{h}^+_l(kr) \,,
\end{equation}
where $N$ is a normalization constant.
An illustration of the analytic structure of the $S$ matrix in terms of $k$
is shown in Fig.~\ref{fig:s_matrix}.

%%%%%%%%%%%%%%%%%%%%%%%%%%%%%%%%%%%%%%%%%%%%%%%%%%%%%%%%%%%%%%%%%%%%%%%%%%%%%%%%%%%%%%%%%
%%%%%%%%%%%%%%%%%%%%%%%%%%%%%%%%%%%%%%%%%%%%%%%%%%%%%%%%%%%%%%%%%%%%%%%%%%%%%%%%%%%%%%%%%

For bound states, the $S$-matrix $s_l(k)$ has corresponding poles on the positive imaginary axis, and it is customary to define the \emph{binding momentum} $\kappa$ by writing $k=\ii\kappa$, with $\kappa>0$.
At such poles, bound-state wave functions are uniquely defined~\cite{faldt97_2542} by $\kappa$ and we can write
\begin{equation}
 \psi_{l,k}(r) \xrightarrow[r \rightarrow \infty]{} N \, \hat{h}^+_l(\ii\kappa r) \,,
\end{equation}
which for an $S$-wave bound state reduces to
\begin{equation}
 \psi_{0,k}(r) \xrightarrow[r \rightarrow \infty]{} N \, \exp({-}\kappa r) \,.
\end{equation}
For all $l$ it is apparent from the behavior of the Riccati-Hankel functions for
positive imaginary arguments that bound-state wave functions tend to zero exponentially
for large $r$.
If a bound state exists close to the scattering threshold, $k=0$, it follows from the
analyticity of the partial-wave $S$ matrix as a function of $E \sim k^2$ that the scattering cross section gets enhanced
at small $k$.

Resonances are another phenomenon that can cause enhancement of the scattering cross section.
While the physically intuitive interpretation of resonances describes them
as short-lived ``metastable'' states, formal scattering theory associates
resonances with complex $S$-matrix poles, leading to an enhancement of the scattering cross section if they are located near the positive real axis (in either momentum or energy representation).
These poles are known as ``Gamow'' or ``Gamow-Siegert'' states.
It is clear that they are not ordinary eigenstates of the Hamiltonian $H = H_0 + V$, which, as a Hermitian operator, can only have real eigenvalues.
There is, however, a well-defined extension of scattering theory to
incorporate resonances as eigenstates in a generalized sense by introducing the
concept of so-called ``rigged Hilbert spaces (RHSs).''
In fact, already scattering states are not normalizable in the sense of possessing square-integrable wave functions, and therefore they do not reside within the ordinary Hilbert space.
From this perspective, they should strictly be considered within the RHS framework.
For practical applications, however, the mathematical complexity associated with this is rarely necessary and can be avoided by, for example, restricting the discussion to the radial Schr\"odinger equation and the properties
of wave functions that solve this ordinary differential equation.\footnote{%
In momentum space, one can obtain a full description of scattering observables
by solving the Lippmann-Schwinger equation for the $T$ matrix.}
In the same spirit, we can characterize decaying Gamow states as solutions of
Eq.~\eqref{eq:SEq-rad} that satisfy the asymptotic boundary condition specified in
Eq.~\eqref{eq:pobc}, albeit with a complex $k$ satisfying $\Rp(k)>0$ and
$\Ip(k)<0$, \ie, located in the \nth{4} quadrant of the complex-momentum
plane.

\subsection{$S$-matrix pole trajectories}
\label{sec:Trajectory}

If the potential supports a bound state, it is possible, as a theoretical exercise, to gradually weaken its strength to move the associated pole into the complex-momentum plane.
The trajectory of the pole depends on the angular momentum of the state and the details of the potential.
For example, the pole associated with an $S$-wave bound state generated by a purely
attractive potential will simply move down the imaginary axis in the complex-momentum plane and become a virtual state after crossing through the $k=0$ threshold.
In the complex-energy plane, the bound-state pole, located on the negative real axis, first moves towards the origin at $E=0$ and then moves backward as a virtual state on the second Riemann sheet of the $S$ matrix.
The two sheets of the $S$ matrix as a function of $E$ are determined by the double-branched nature of the square-root function,
$k = \pm\sqrt{2\mu E}$.
The standard convention, which we also adopt here, is to attach the two sheets along the positive real axis, the so-called ``unitarity cut.''

\begin{figure}[htb]
    \centering
    \includegraphics[width=0.49\textwidth]{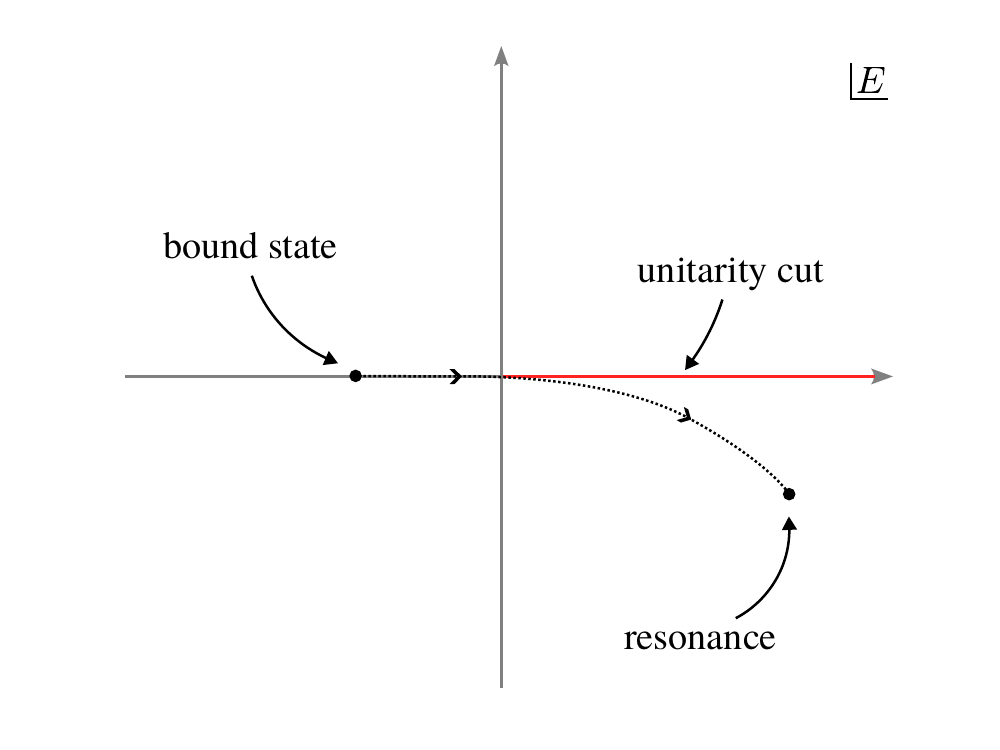}
    \caption{An illustration of a bound state moving through the complex-energy plane
    as the potential is made gradually weaker. In this case, it crosses the threshold
    and moves into the second Riemann sheet to
    become a resonance (Gamow state) with a complex energy.}
    \label{fig:crossing}
\end{figure}

If instead the potential has a barrier, either from the actual shape of $V(r)$ or \emph{effectively} due to the nonzero centrifugal term in Eq.~\eqref{eq:SEq-rad} for $l>0$, the pole may also move through the threshold into the \nth{4} quadrant (in both the momentum and the energy planes) so that the state becomes a (decaying) resonance.
This is the scenario that is of primary interest to us in this work.
In particular, in Sec.~\ref{sec:B2R} we develop a strategy to \emph{extrapolate}
along such a trajectory, as illustrated in Fig.~\ref{fig:crossing}.

%%%%%%%%%%%%%%%%%%%%%%%%%%%%%%%%%%%%%%%%%%%%%%%%%%%%%%%%%%%%%%%%%%%%%%%%%%%%%%%%%%%%%%%%%
%%%%%%%%%%%%%%%%%%%%%%%%%%%%%%%%%%%%%%%%%%%%%%%%%%%%%%%%%%%%%%%%%%%%%%%%%%%%%%%%%%%%%%%%%

\subsection{Complex-scaling method (CSM)}
\label{sec:CS}

As described above, bound states have associated imaginary momenta $k=\ii\kappa$ with
real $\kappa>0$, whereas resonances are described by complex $k$ with $\Ip(k)<0$.
This means that asymptotically, resonance wave functions grow exponentially with $r$
and are therefore---like scattering states but in some sense even more so---not
square-integrable; \ie, they do not correspond to normalizable states in the ordinary
Hilbert space.
While the rigged Hilbert space construction offers a rigorous mathematical formalism
to deal with this difficulty (see, for example,
Ref.~\cite{delaMadrid:2012aa} for an introduction), for practical calculations there exists a much simpler
alternative.
The so-called (uniform) \emph{complex-scaling method}~\cite{reinhardt82_708,Moiseyev:1998aa} enables a description of
resonances with, essentially, bound-state techniques.
This is achieved by expressing the wave function not as usual along the real $r$ axis, but
on a contour rotated into the complex-$r$ plane.
This can be achieved by applying the transformation
\begin{equation}
 r \to r e^{\ii \phi}
\label{eq:r-scaled}
\end{equation}
to Eq.~\eqref{eq:SEq-rad}, with some angle $\phi$.
The proper choice of $\phi$ in general depends on the position of the
resonance one wishes to study.
If the state of interest has a complex energy $E$, then it is necessary to ensure
that $\phi > {-}\frac{\arg{E_r}}{2}$.
As $E$ is usually not known beforehand, one might repeat
the calculation while increasing $\phi$ until a resonance is found.\footnote{%
One might think of simply setting $\phi=\pi/4$ to accommodate all possible resonances.
However, in most cases, large $\phi$ angles lead to potentials and wave functions not vanishing fast enough along the contour,
thereby demanding more expensive calculations.}

\begin{figure}
    \centering
    \includegraphics[width=0.49\textwidth]{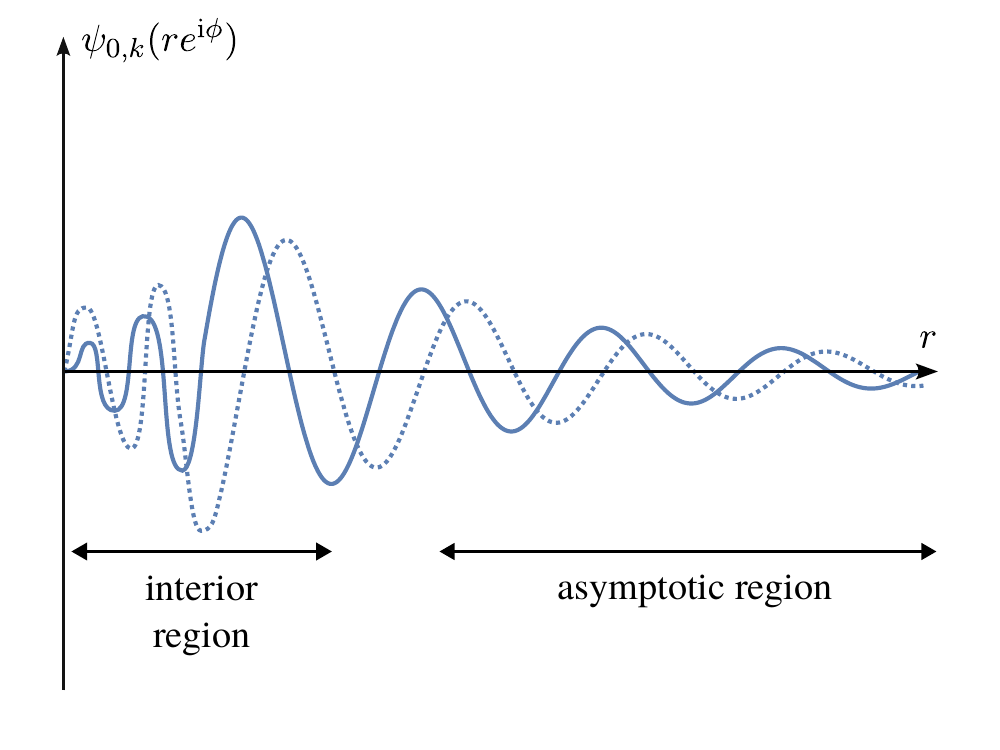}
    \caption{Illustration of the reduced radial wave function of a typical $S$-wave resonance (Gamow state),
    on a complex-scaled $r$ contour. The solid (dotted) line corresponds to the real
    (imaginary) part. It asymptotically converges to the Riccati-Hankel
    function $\hat{h}^+_0(kr e^{\ii \phi})=\hat{h}^+_0(\Tilde{k}r)=\exp(\ii \Tilde{k} r)$,
    where we define $\Tilde{k} = k e^{\ii \phi}$, the \emph{effective} wave number with $\Ip(\Tilde{k})>0$,
    so that it is normalizable just like bound-state wave functions.}
    \label{fig:wavefunc}
\end{figure}

With the convention in Eq.~\eqref{eq:r-scaled}, $r$ is still a real parameter but no longer describes
the physical radial coordinate of the system.
The overall argument $kr e^{\ii \phi}$ of the Riccati-Hankel function in Eq.~\eqref{eq:pobc} satisfies
$\Ip(kr e^{\ii \phi})>0$, and therefore square-integrability of the wave function as a function of
$r$ is recovered.
An example of such a scaled wave function is illustrated in Fig.~\ref{fig:wavefunc}.

%%%%%%%%%%%%%%%%%%%%%%%%%%%%%%%%%%%%%%%%%%%%%%%%%%%%%%%%%%%%%%%%%%%%%%%%%%%%%%%%%%%%%%%%%
%%%%%%%%%%%%%%%%%%%%%%%%%%%%%%%%%%%%%%%%%%%%%%%%%%%%%%%%%%%%%%%%%%%%%%%%%%%%%%%%%%%%%%%%%

It was shown in Ref.~\cite{Afnan:1991kb} that the scaling of the radial coordinate $r$ is
equivalent to a rotation in momentum representation that goes in the opposite (clockwise)
direction with the same angle $\phi$.
That is, if we consider the wave function of the state as a function of a momentum coordinate
$q$, then complex scaling is implemented via
\begin{equation}
 q \to q e^{{-}\ii \phi} \,.
\end{equation}
This procedure then makes it possible to alternatively calculate resonance wave
functions in momentum space.
Note furthermore that scaling in momentum space can also be understood
as a rotation
of the branch cut in the complex-energy plane by an angle $2\phi$ clockwise, thereby
exposing a section of the second Riemann sheet where resonances are located.

\begin{figure}[htb]
    \centering
    \begin{overpic}[width=0.49\textwidth]{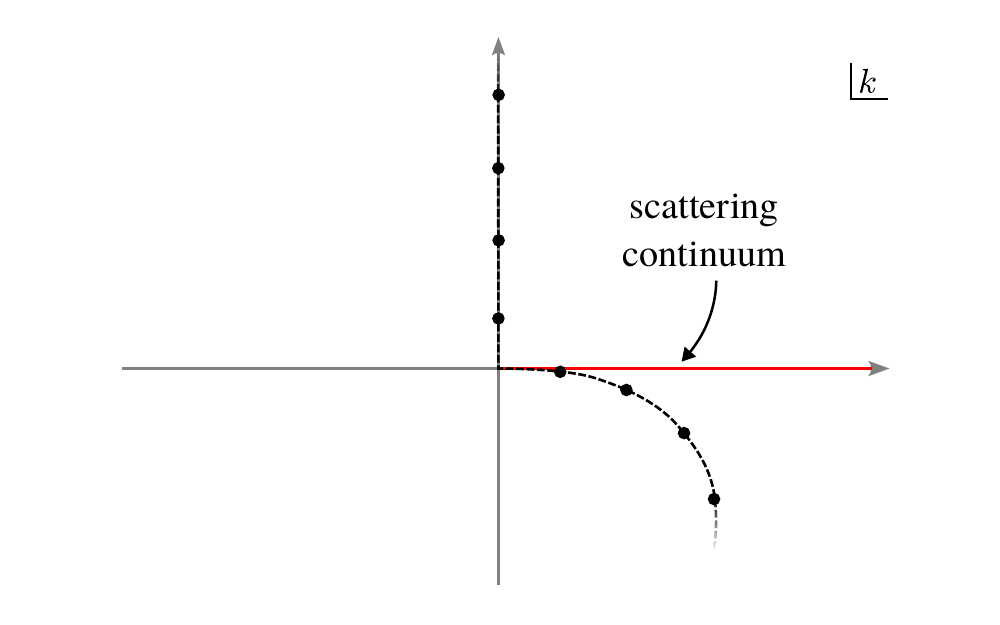}
        \put(6,4){(a)}
    \end{overpic} \\
    \begin{overpic}[width=0.49\textwidth]{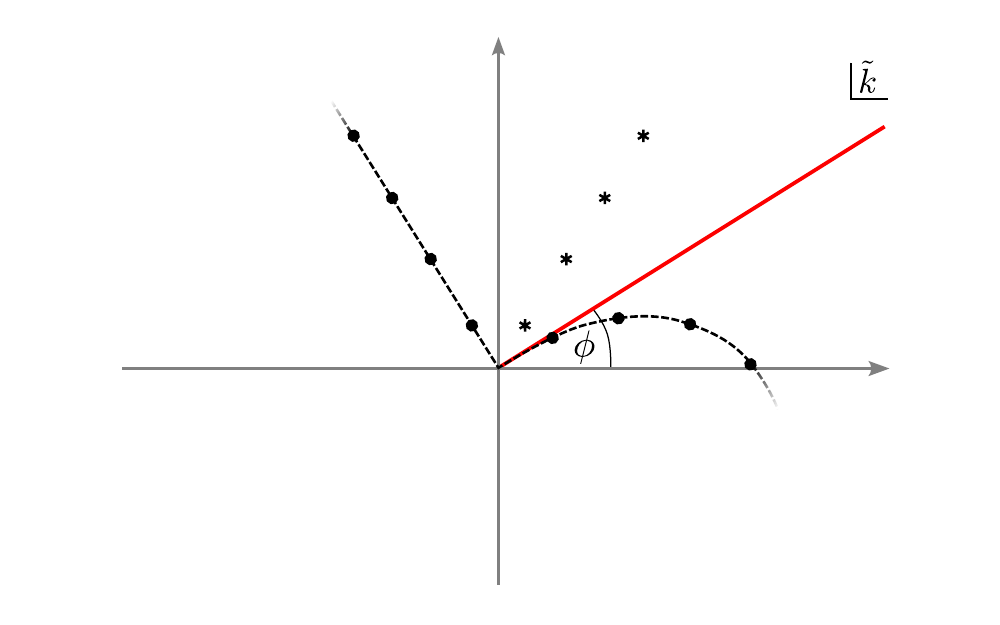}
        \put(6,4){(b)}
    \end{overpic}
    \caption{Trajectory of an $S$-matrix pole in the $k$ plane before (a) and
    after (b) complex scaling. Thanks to the CSM, resonance solutions have
    $\Ip(\Tilde{k})>0$ and thus are square-integrable.
    Also note how complex-scaled bound states come with ``negative" wave numbers [$\Rp(\Tilde{k})<0$]
    while resonances come with ``positive" wave numbers [$\Rp(\Tilde{k})>0$].
    Indicated by asterisks are the CA-EC vectors discussed in Sec.~\ref{sec:B2R}.
    They are obtained by complex conjugation of
    bound-state wave functions and will have asymptotic wave numbers
    ${-}\Tilde{k}^*=\Tilde{k} e^{{-}2\ii \phi}$, placing them in the 1st quadrant and
    closer to the resonance regime.}
    \label{fig:rotation}
\end{figure}

After this transformation, we can absorb the $e^{\ii \phi}$ phase into the wave number $k$
and define the \emph{effective} wave number as $\Tilde{k} = k e^{\ii \phi}$,
so that the asymptotic form in Eq.~\eqref{eq:pobc} is preserved as
\begin{equation}
 \psi_{l,k}(r e^{\ii \phi}) \xrightarrow[r \rightarrow \infty]{} N \, \hat{h}^+_l(\Tilde{k} r) \,.
\end{equation}
The scaling technique can thus be interpreted as mapping resonances from the
\nth{4} quadrant in the complex-$k$ plane to the \nth{1} quadrant in the $\Tilde{k}$ plane (see Fig.~\ref{fig:rotation}).
For future reference, we note that, at the same time, it will effectively map bound states
from the positive imaginary $k$ axis to the \nth{2} quadrant in the complex $\Tilde{k}$ plane.

%%%%%%%%%%%%%%%%%%%%%%%%%%%%%%%%%%%%%%%%%%%%%%%%%%%%%%%%%%%%%%%%%%%%%%%%%%%%%%%%%%%%%%%%%
%%%%%%%%%%%%%%%%%%%%%%%%%%%%%%%%%%%%%%%%%%%%%%%%%%%%%%%%%%%%%%%%%%%%%%%%%%%%%%%%%%%%%%%%%

\subsection{Non-Hermiticity and the c-product}

In traditional quantum mechanics, one requires the Hamiltonian $H$ to be Hermitian ($H^\dagger=H$)
to ensure that the energy spectrum, being a physical observable, is real and that time evolution is strictly unitary, \ie, the norm of quantum states are preserved under the time
evolution operator $\ee^{{-}\ii H t / \hbar}$.
However, when considering decay, an inherently time-dependent phenomenon, in a time-independent framework such as the complex-scaling method, the Hamiltonian is no longer Hermitian.
Instead, in the present case, it becomes complex symmetric
($H^T=H$)~\cite{Moiseyev:1998aa}.
This permits the energy spectrum to include complex eigenvalues, which, as discussed in Sec.~\ref{sec:Basics}, is precisely what is needed to describe resonances.
In fact, the non-Hermiticity and the corresponding nonunitary time evolution of Gamow states are well aligned with the physical interpretation of resonances as metastable systems that ultimately decay.

Similarly to how nondegenerate eigenvectors of a Hermitian operator are orthogonal under
the inner product defined on the Hilbert space, the nondegenerate eigenvectors of a complex
symmetric operator are  orthogonal under the so-called
``c-product''~\cite{Moiseyev:1978aa,Moiseyev:2011}.
For eigenstates $\ket{\psi_1}$ and $\ket{\psi_2}$ with equal angular-momentum
quantum numbers $(l,m)$, we define the c-product in coordinate representation as
\begin{equation}
 \braket{\psi_1 | \psi_2} = \int \dd r \, \psi_1(r)\psi_2(r) \,,
\label{eq:CP-vecr}
\end{equation}
and similarly in momentum space.
Note that $\psi_1(r)$ appears without complex conjugation under the integral.
This is precisely the c-product introduced in Ref.~\cite{Moiseyev:1978aa} with the
notation $(\psi_1|\psi_2)$.
In this paper, we use the standard notation $\braket{\psi_1 | \psi_2}$ with the implicit
understanding that for complex-scaled systems this is meant to denote the c-product.

Equivalently, one can change the definition of bra states so that no complex conjugation is involved
when they are associated with a complex-scaled system.
This is so even for bound states calculated with complex scaling.
Although the energies of such states remain
real, wave functions become complex when defined along the rotated contour and the orthogonality of states
with different binding energies is ensured only if no complex conjugation is performed for bras, leading
again to the c-product~\cite{Moiseyev:2011}.
Ultimately, these concepts can be understood by properly distinguishing bra and ket states as, respectively, left and right
eigenvectors of the non-Hermitian complex-scaled Hamiltonian~\cite{Afnan:1991kb}.
Even more rigorously, a comprehensive theory for Gamow bras and kets can be developed within the RHS formalism mentioned previously~\cite{delaMadrid:2012aa}.
However, in practice we find it convenient and sufficient to employ complex scaling along with the
c-product.

\section{Resonance continuation}
\label{sec:Results}

We now discuss the extension of eigenvector continuation to resonance states.
Generally, EC works by obtaining eigenstates of a Hamiltonian $H(c)$ with a parametric
dependence on a parameter $c$ for several values of that parameter.\footnote{%
For simplicity, we assume here that there is only one scalar parameter and note that the
extension to multiple parameters is straightforward~\cite{Konig:2019adq}.}
The set of parameters $\{c_i\}$ used for this step is referred to as ``training points,''
and the corresponding ``training vectors'' $\ket{\psi(c_i)}$ are used to construct an
effective basis within which the problem is subsequently solved for one or more target values
of the parameter $c$.
For typical applications of EC, this procedure reduces the dimension of the
problem from a large Hilbert space to the small subspace spanned by the training vectors, thereby
leading to a vast reduction of the computational cost for each target evaluation.
Specifically, if we denote the target point as $c_*$, EC involves solving the generalized
eigenvalue problem
\begin{equation}
 H_{\text{EC}} \ket{\psi(c_*)}_{\text{EC}}
 = E(c_*)_{\text{EC}} \, N_{\text{EC}} \ket{\psi(c_*)}_{\text{EC}} \,,
\label{eq:EC-GEVP}
\end{equation}
with the following Hamiltonian and norm matrices:
\begin{align}
\label{eq:EC-H}
 \big(H_{\text{EC}}\big)_{ij} &= \braket{\psi(c_i)|H(c_*)|\psi(c_j)} \,, \\
\label{eq:EC-N}
 \big(N_{\text{EC}}\big)_{ij} &= \braket{\psi(c_i) | \psi(c_j)} \,.
\end{align}
The key to making this remarkably simple prescription useful is that typically EC is able to
construct highly effective variational bases, with rapid convergence as the number of training data is
increased~\cite{Sarkar:2020mad}.

Eigenvector continuation involving resonance states can be defined by prescribing that
the matrix elements in Eqs.~\eqref{eq:EC-H} and~\eqref{eq:EC-N} are to be evaluated
using the c-product.
Importantly, this is to be used in connection with the complex-scaling technique described
in Sec.~\ref{sec:CS}, so that for evaluating the c-product one integrates along the rotated
contour in either $r$ space or $q$ space.
This procedure ensures in particular that all matrix elements remain well defined and
\emph{finite}, which would not be the case without complex scaling because, as
previously mentioned, Gamow states would then not be normalizable and their wave functions
would exhibit, in coordinate representation, an exponentially growing amplitude.
The Hamiltonian and norm matrices, $H_{\text{EC}}$ and $N_{\text{EC}}$, obtained with
the c-product will not be Hermitian but complex symmetric, and therefore they may have
complex eigenvalues, as is required to describe resonances.

\paragraph*{Numerical tests.}

We can show with explicit
examples that indeed this procedure works nicely in practice.
All of the calculations shown in the following sections were performed using a discrete momentum basis
with a cutoff of $\Lambda=8.0$ (in the dimensionless units explained above) and consisting of $N = 256$ mesh points distributed
according to a 256-point Gauss–Legendre quadrature to ensure convergence.
In general, the proper choice of $\Lambda$ and $N$ depends on the properties of $V$, and we have checked that the above is sufficient to ensure numerical
convergence for the particular examples we describe below.
Furthermore, the momentum basis is complex-scaled as discussed previously, by an angle $\phi=\pi/6$.
As a crude way to estimate the uncertainty of the EC extrapolations,
we repeat every calculation $128$ times while randomizing the training
points within the given interval.
Finally, the distributions of the extrapolation
results are indicated in figures by their $68.2\%$ and $95.4\%$ percentile intervals,
the approximate intervals corresponding to 1 and 2 standard deviations, respectively.

All of the potentials we use for numerical tests are based on a local Gaussian form,
which in configuration space reads
\begin{equation}
    F(\alpha,r)=\exp\left(-\alpha r^2\right) \,.
\end{equation}
For momentum-space calculations, we take the Hankel transform of the above.
This is given by
\begin{equation}
    F_l(\alpha,q,q')=\frac{2}{\pi} \int_0^\infty \dd r \,
    \hat{j}_l(qr) \exp\left(-\alpha r^2\right) \hat{j}_l(q'r) \,,
\end{equation}
where $\hat{j}_l(z)$ are the Riccati-Bessel functions (see Ref.~\cite[p.~182]{Taylor:1972pty}).
For $l=0$, this reduces to
\begin{equation}
    F_0(\alpha,q,q')=\frac{1}{\sqrt{\alpha\pi}} \exp\left(-\frac{q^2+q'^2}{4\alpha}\right)
    \sinh\left(\frac{qq'}{2\alpha}\right) \,.
\end{equation}
For $l=1$, we have
\begin{spliteq}
    F_1(\alpha,q,q')=\sqrt{\frac{\alpha}{4\pi}} \left[ \left(\frac{1}{\alpha}+\frac{2}{q q'}\right) \exp\left(-\frac{(q+q')^2}{4\alpha}\right) \right. \\
    \left. + \left(\frac{1}{\alpha}-\frac{2}{q q'}\right) \exp\left(-\frac{(q-q')^2}{4\alpha}\right) \right] \,.
\end{spliteq}
Because there is no centrifugal term in the $S$-wave radial equation, we include in the $l=0$ potential a
repulsive barrier to support resonances:
\begin{equation}
    \label{eq:V_for_s_wave}
    V(c,q,q')=c\left[-5 F_0\left(\nicefrac{1}{3},q,q'\right)
    +2 F_0\left(\nicefrac{1}{10},q,q'\right)\right] \,.
\end{equation}
For a $P$-wave example, the following simple Gaussian potential is considered:
\begin{equation}
    \label{eq:V_for_p_wave}
    V(c,q,q')=-c F_1\left(\nicefrac{1}{4},q,q'\right) \,.
\end{equation}
Although not exploited here, we note that Hamiltonians $H(c)$ with a simple
linear dependence on $c$ (like the ones we consider), or more generally, an
affine dependence on a vector of parameters $\vec{c}$, permit further
optimization of the EC calculation via decomposition into off-line and on-line
tasks, see for example Ref.~\cite{Drischler:2022ipa}.

As Supplemental Material, we provide the code for our calculations as
downloadable files~\cite{supp}.
The setup is split into a \textsc{python} library, \verb|twobodyEC.py|, that implements
the basic numerical techniques discussed above, and a \textsc{jupyter} notebook,
\verb|calculations.ipynb|, that reproduces the exact numerical examples 
presented in the following.

\subsection{Resonance-to-resonance extrapolation}
\label{sec:R2R}

We now consider a Hamiltonian $H(c)$ that supports a resonance for some range of $c$
and we implement the standard EC prescription by first constructing $H(c_i)$ on a
complex-scaled basis for several training points $\{c_i\}$.
In our calculation, this amounts to calculating the matrix elements
$\braket{q_n e^{{-}\ii\phi} | H(c_i) | q_m e^{{-}\ii\phi}}$ using a $q$ momentum mesh
as described previously (with $\phi>{-}\frac{\arg{E_r}}{2}$, where $E_r$ is the complex energy associated with
the resonance for all $c$ within the region of interest).
Then we proceed to solve the system to obtain
the exact eigenvectors $\ket{\psi(c_i)}$ for each $c_i$.
This is the part that takes the bulk of the computational time.

We now want to determine $E(c_*)$ at some target point $c=c_*$ via extrapolation.
To that end, we construct $H(c_*)$, but instead of determining its eigenvalues directly, we project
$H(c_*)$ onto the ``EC subspace'' spanned by $\{\ket{\psi(c_i)}\}$.
In practice, this is done by calculating
the projected matrix elements $(H_\text{EC})_{i,j} = \braket{\psi(c_i) | H(c_*) | \psi(c_j)}$ and
the norm matrix elements $(N_\text{EC})_{i,j} =\braket{\psi(c_i) | \psi(c_j)}$.
The latter accounts for the nonorthogonality of the basis vectors.
The c-product prescription has to be followed in this step.
Finally, we diagonalize the much smaller projected matrix
$H_\text{EC}$ by solving the generalized eigenvalue problem\footnote{%
Alternatively, one could orthonormalize $\{\ket{\psi(c_i)}\}$ beforehand,
again following the c-product formalism, and then solve an ordinary
eigenvalue problem.}
\begin{equation}
    \label{eq:GEVP}
    H_\text{EC} \, \ket{\psi(c_*)}_\text{EC} = E(c_*)_\text{EC} \, N_\text{EC} \, \ket{\psi(c_*)}_\text{EC} \,.
\end{equation}
Equation~\eqref{eq:GEVP} yields a spectrum of complex eigenvalues that includes the approximation $E(c_*)_\text{EC}$
corresponding to the exact energy eigenvalue $E(c_*)$ of the particular state we are interested in.
Identification of the relevant $E(c_*)_\text{EC}$
for non-Hermitian systems cannot be based on the variational principle, and so,
unlike EC applied to bound states, it is not sufficient here to simply pick extremal eigenvalues from the EC spectrum.
Because for the benchmark presented here the exact value $E(c_*)$ is calculated alongside the extrapolated spectrum, we employ the criterion $\min|E(c_*)_\text{EC}-E(c_*)|$ to identify the proper state within the complex spectrum.
For practical applications of the technique, the exact $E(c_*)$ will
typically be unknown.
To deal with this situation, one might manually follow the extrapolated pole as it gradually crosses the threshold from being a bound state (for which the identification is straightforward) to becoming a
resonance.
More generally, one may employ a systematic overlap-based technique as is common practice in the calculation of resonances within the Berggren basis~\cite{michel03_10}.

\begin{figure}[htb]
    \centering
    \includegraphics[width=0.49\textwidth]{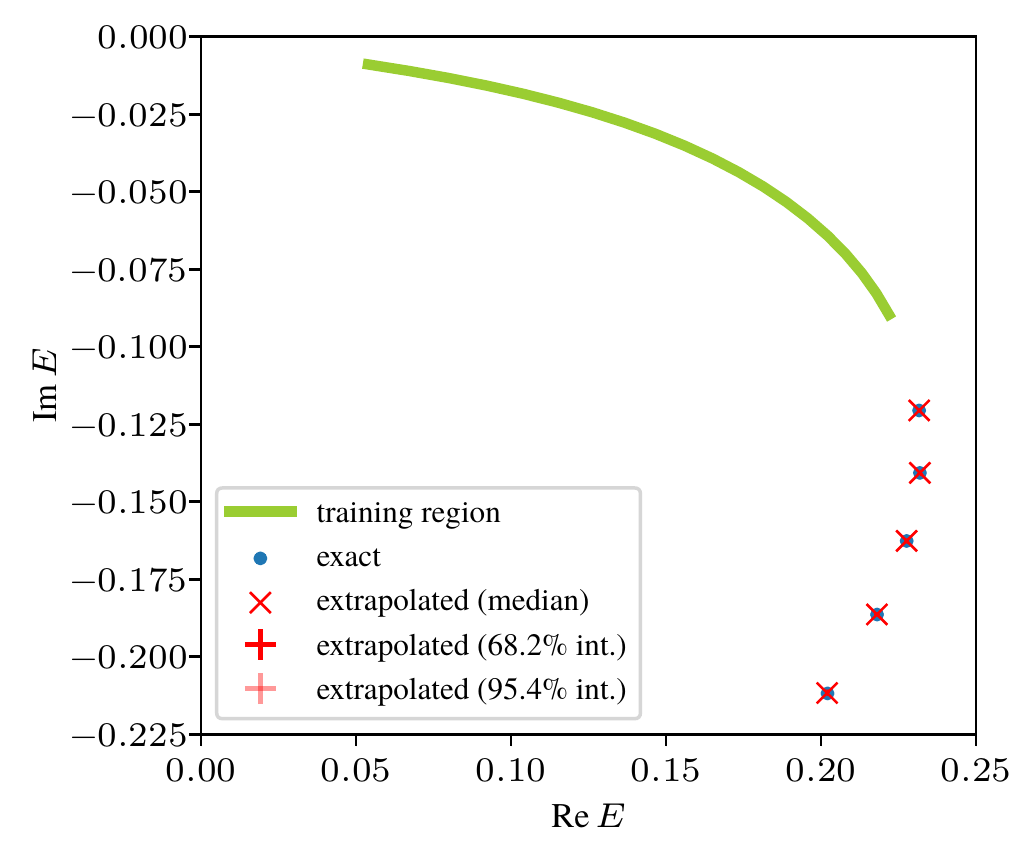}
    \caption{Application of EC for resonance-to-resonance extrapolation.
    Five training points were randomly drawn from the region
    $c\in (0.45,0.78)$ per dataset.
    Apart from using the c-product formalism,
    the calculation proceeds similar to ordinary EC, yielding accurate approximations for the
    complex-energy eigenvalues of resonances (Gamow states). See text for details.}
    \label{fig:R2R}
\end{figure}

To exemplify resonance-to-resonance continuation, the system described by
Eq.~\eqref{eq:V_for_s_wave} is considered.
Results are shown in Fig.~\ref{fig:R2R}.

%%%%%%%%%%%%%%%%%%%%%%%%%%%%%%%%%%%%%%%%%%%%%%%%%%%%%%%%%%%%%%%%%%%%%%%%%%%%%%%%%%%%%%%%%
%%%%%%%%%%%%%%%%%%%%%%%%%%%%%%%%%%%%%%%%%%%%%%%%%%%%%%%%%%%%%%%%%%%%%%%%%%%%%%%%%%%%%%%%%

\subsection{Bound-state-to-resonance extrapolation}
\label{sec:B2R}

While applying EC solely within the resonant regime is interesting and certainly useful
in practice (\eg to produce EC-based emulators for
uncertainty quantification~\cite{Konig:2019adq}), it is a more fascinating question
whether we can set up an extrapolation scheme that uses training vectors
at $c$ values corresponding only to \emph{bound states}, but which then extrapolates
to $c_*$ where the state is a resonance.
In other words, we would like to use EC to extrapolate along the $S$-matrix pole
trajectories described in Sec.~\ref{sec:Trajectory} from the regime of bound states into
the resonance domain.
We note that, in general, not all bound states transition into resonances as illustrated in
Sec.~\ref{sec:Trajectory}, but in order to illustrate the method we consider
here only cases where it is known \apriori that this is the case.

Clearly, a naive approach without appropriate complex scaling of the basis will not be successful in predicting resonance
energies because $H_{\text{EC}}$ and $N_{\text{EC}}$ will be trivially Hermitian
with that prescription.
However, even with complex scaling and the matrix elements defined in terms of the c-product, it is not
possible to obtain complex energies via EC because $H_{\text{EC}}$ and $N_{\text{EC}}$ will, in fact, be real and symmetric.
This can be seen as follows.
If we use the notation $\psi(c_i;r)$ for the (reduced) radial
wave function corresponding to the state $\ket{\psi(c_i)}$,
then for the norm matrix it holds that
\begin{spliteq}
\label{eq:N_real}
 (N_{\text{EC}})_{ij} &=\braket{\psi(c_i) | \psi(c_j)} \\
  &=\int_C \dd r \, \psi(c_i;r) \psi(c_j;r) \\
  &=\int_0^\infty \dd r \, \psi(c_i;r) \psi(c_j;r) \in\RR \,,
\end{spliteq}
where $\int_C$ denotes integration along the complex-scaled contour.
We have used the fact that the contour can be rotated back to the real axis without changing the value of the integral because no singularities are swept over and for bound states the contribution of the arc at infinity that closes the curve between the real axis and the rotated contour vanishes.\footnote{%
The bound-state wave functions remain strictly
normalizable even with the $r$ contour rotated into the upper half-plane.}
Then, we have made use of the fact that ``c-normalized'' bound-state wave functions are real along the real axis.
This is so because any bound-state wave function $\varphi(r)$ can be chosen to be real and any arbitrary global factor $N$ will be constrained to
$N\in\RR$ by the c-normalization condition
\begin{equation}
    \int_0^\infty \dd r \, \left[N\varphi(r)\right]^2 = 1 \,,
\end{equation}
resulting in a real normalized wave function $\psi(r)=N\varphi(r)$.
$N_{\text{EC}}$ is also trivially symmetric due to the properties of the c-product.

By the same token, for the projected Hamiltonian matrix, we see that
\begin{spliteq}
\label{eq:H_real}
 (H_{\text{EC}})_{ij} &=\braket{\psi(c_i) | H(c_*) | \psi(c_j)} \\
  &=\int_C \dd r \int_C \dd r' \, \psi(c_i;r) H(c_*;r,r') \psi(c_j;r') \\
  &=\int_0^\infty \dd r \int_0^\infty \dd r' \, \psi(c_i;r) H(c_*;r,r') \psi(c_j;r') \\
  & \in\RR \,,
\end{spliteq}
noting that everything under the final integral is real.
This shows that $H_{\text{EC}}$ has only real entries.
Because $H(c_*)$ is complex only due to the contour rotation, by the ``turn over
rule''~\cite{Moiseyev:1998aa} it follows moreover that $H_{\text{EC}}$ is
symmetric, which concludes the proof.

\begin{figure}[htb]
    \centering
    \includegraphics[width=0.49\textwidth]{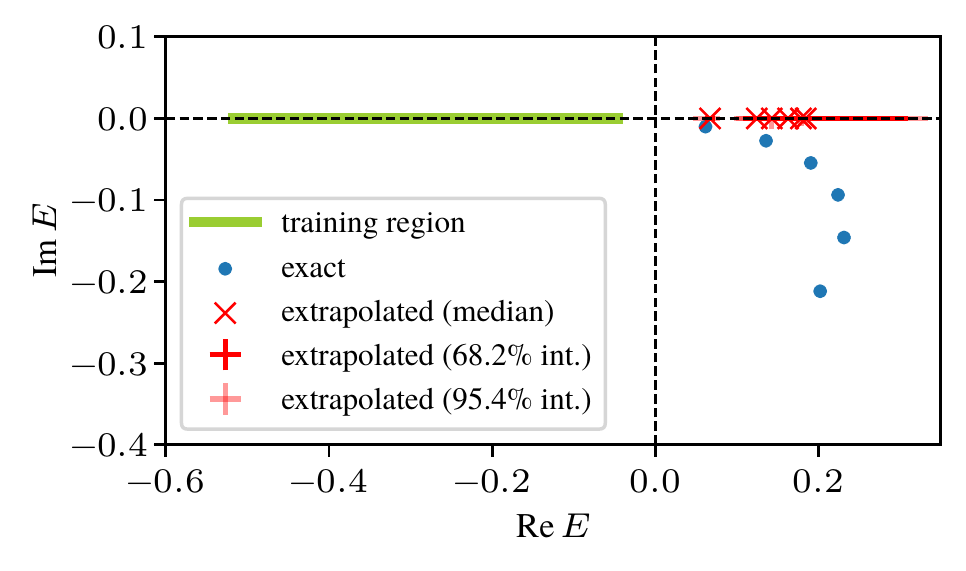}
    \caption{Attempt to extrapolate from bound states to resonances using ordinary EC
    without augmentation.
    Five training points were randomly drawn from the region $c\in (0.9,1.3)$ per dataset.
    See text for details.}
    \label{fig:noCAEC}
\end{figure}

The same $S$-wave potential given in Eq.~\eqref{eq:V_for_s_wave} is considered
for demonstrating the failure of a naive bound-state-to-resonance extrapolation.
As shown in Fig.~\ref{fig:noCAEC}, the extrapolated energies do not extend
beyond the real axis, as one would expect for resonance states.

%%%%%%%%%%%%%%%%%%%%%%%%%%%%%%%%%%%%%%%%%%%%%%%%%%%%%%%%%%%%%%%%%%%%%%%%%%%%%%%%%%%%%%%%%
%%%%%%%%%%%%%%%%%%%%%%%%%%%%%%%%%%%%%%%%%%%%%%%%%%%%%%%%%%%%%%%%%%%%%%%%%%%%%%%%%%%%%%%%%

\paragraph*{Conjugate-augmented EC.}

Fortunately, it turns out that there is a way to accomplish bound-state-to-resonance
extrapolations with an extension of the EC prescription.
The appropriate strategy, which we refer to as ``conjugate-augmented eigenvector
continuation (CA-EC)'' and which we elucidate further below, is to enlarge the
subspace spanned by the bound-state training vectors by including, \emph{in addition},
complex-conjugate versions of the existing wave functions.
This can be easily implemented numerically (in any concrete representation of the wave
functions) by duplicating training vectors stored in memory with elementwise
complex conjugation.\footnote{%
Note that the extra memory cost for storing the additional vectors can be avoided
if one performs the complex conjugation for the extra states on the fly during the
construction of $H_{\text{EC}}$ and $N_{\text{EC}}$.}

As we show with concrete examples below, CA-EC works
nicely in practice.
To understand why that is so, note that after complex scaling the asymptotic wave
functions (as functions of $r$ along the rotated contour) will have decaying
wave numbers for both bound states and resonances, \ie, $\Ip(\Tilde{k})>0$, as
illustrated in Fig.~\ref{fig:rotation}.
However, $\Rp(\Tilde{k})$ is negative for bound states and positive for resonances.
EC is ineffective at extrapolating complex ``plane waves'' of the form
$\hat{h}^+_l(\Tilde{k}r)$ with rapidly changing wave numbers, especially when the sign
of the real part is supposed to change upon extrapolating from the training regime to the
target point.
This systematic deficiency of the basis can be remedied by including additional
vectors that have ``positive'' asymptotic wave numbers, \ie, $\Rp(\Tilde{k})>0$.
Exactly this is achieved with CA-EC because the complex-conjugated wave functions
have such asymptotic wave numbers.
Specifically, the asymptotic wave number of a complex-conjugated bound state will be
${-}\Tilde{k}^*=\Tilde{k} e^{{-}2\ii \phi}$ if $\Tilde{k}$ denotes the wave number of
the bound state (\cf~Fig.~\ref{fig:rotation}).

\begin{figure}[htb]
    \centering
    \includegraphics[width=0.49\textwidth]{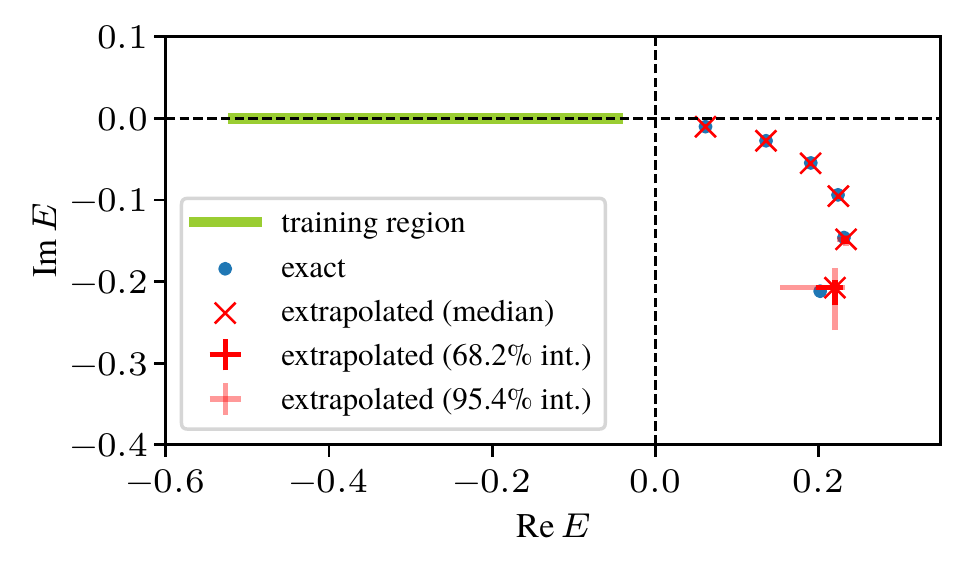}
    \caption{Bound-state-to-resonance extrapolation performed with CA-EC for the system given by Eq.~\eqref{eq:V_for_s_wave}.
    Five training points were randomly drawn from the region $c\in (0.9,1.3)$ per dataset.
    See text for details.}
    \label{fig:CAEC}
\end{figure}

\begin{figure}[htb]
    \centering
    \includegraphics[width=0.49\textwidth]{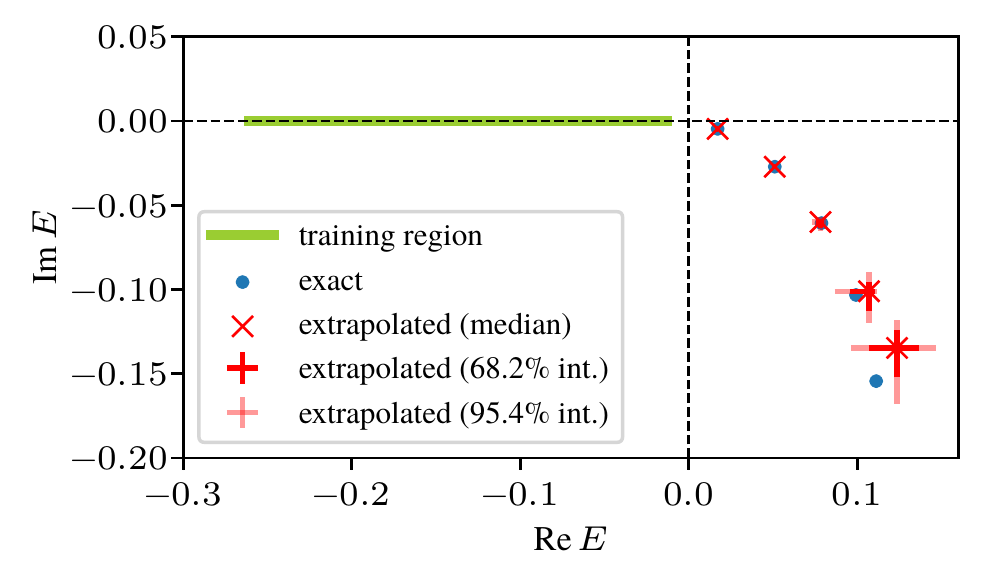}
    \caption{Bound-state-to-resonance extrapolation with CA-EC for the Gaussian potential
    shown in Eq.~\eqref{eq:V_for_p_wave} in the $P$ wave.
    Five training points were randomly drawn from the region $c\in (3.1,4.0)$ per dataset.
    See text for details.}
    \label{fig:B2R_p_wave}
\end{figure}

The systems described by Eqs.~\eqref{eq:V_for_s_wave} and~\eqref{eq:V_for_p_wave} are considered
to illustrate the CA-EC bound-state-to-resonance extrapolation method.
As shown in Figs.~\ref{fig:CAEC} and~\ref{fig:B2R_p_wave}, CA-EC \emph{can} reproduce resonance states, and the extrapolated energies agree nicely with
exact calculations performed for comparison.

%%%%%%%%%%%%%%%%%%%%%%%%%%%%%%%%%%%%%%%%%%%%%%%%%%%%%%%%%%%%%%%%%%%%%%%%%%%%%%%%%%%%%%%%%
%%%%%%%%%%%%%%%%%%%%%%%%%%%%%%%%%%%%%%%%%%%%%%%%%%%%%%%%%%%%%%%%%%%%%%%%%%%%%%%%%%%%%%%%%

%%%%%%%%%%%%%%%%%%%%%%%%%%%%%%%%%%%%%%%%%%%%%%%%%%%%%%%%%%%%%%%%%%%%%%%%%%%%%%%%%%%%%%%%%
%%%%%%%%%%%%%%%%%%%%%%%%%%%%%%%%%%%%%%%%%%%%%%%%%%%%%%%%%%%%%%%%%%%%%%%%%%%%%%%%%%%%%%%%%

The key to understanding why CA-EC works is the insight that in the interior region (\ie, $r$ sufficiently small), resonance wave functions look similar to those of bound
states.
As shown in Fig.~\ref{fig:wavefunc}, the oscillating and exponentially growing behavior only sets in at larger $r$.
Therefore, in the interior region, an EC basis comprised of bound states \emph{can}
properly express the behavior of the resonance wave function at the target point, and
it is only the asymptotic behavior that needs an enlarged basis to
be properly represented.
To further elucidate this explanation, we can consider an alternative approach where
we augment the original EC basis not with complex-conjugated versions of the training bound-state wave functions, but with Riccati-Hankel functions
$\hat{h}^+_l(\Tilde{k}r)$ that have the same wave numbers $\Tilde{k}$ that are
otherwise provided by the complex-conjugated bound states with CA-EC (\ie asterisks in Fig.~\ref{fig:rotation}).
This approach is somewhat similar to the construction of the so-called
Berggren basis~\cite{Berggren:1968zz,Berggren:1993zz}.

This basis augmentation with Riccati-Hankel functions can be performed in
configuration space as well as in momentum space.
In configuration space, we can use the explicit representation~\cite[Eq.~10.49.6]{DLMF}
\begin{equation}
 \hat{h}^+_l(kr)
 = \exp(\ii kr)\sum_{n=0}^l{\frac{\ii^{n-l-1}}{2^n}
 \frac{(l+n)!}{n!(l-n)!} \frac{1}{(kr)^n}} \,,
\label{eq:RH-r}
\end{equation}
for a state with angular momentum $l$.
In momentum space, we need the Hankel transform of $\hat{h}^+_l(kr)$, which is given
by
\begin{equation}
 \phi_k(q) = \sqrt{\frac{2}{\pi}}\frac{k\left(q/k\right)^l}{q^2-k^2} \,.
\label{eq:RHAEC_mom_space}
\end{equation}
This can be verified by explicitly carrying out the inverse Hankel transform as follows:
\begin{widetext}
\begin{spliteq}
 \sqrt{\frac{2}{\pi}}\int_0^\infty \dd q \, \hat{j}_l(qr) \left[
  \sqrt{\frac{2}{\pi}}\frac{k\left(q/k\right)^l}{q^2-k^2}
 \right]
 &= \frac{2 k}{\pi}\int_0^\infty \dd q \, \hat{j}_l(qr)\,\frac{\left(q/k\right)^{l}}{q^2-k^2} \\
 &= \frac{k}{\ii\pi}\int_0^\infty \dd q \,\left[
  \hat{h}^+_l(qr)-(-1)^l \hat{h}^+_l(-qr)
 \right] \frac{\left(q/k\right)^{l}}{q^2-k^2} \\
 &= \frac{k}{\ii\pi} \left[
  \int_0^\infty \dd q \, \hat{h}^+_l(qr)\frac{\left(q/k\right)^{l}}{q^2-k^2}
  - \int_0^\infty \dd q \, \hat{h}^+_l(-qr)\frac{\left(-q/k\right)^{l}}{q^2-k^2}
 \right] \\
 &= \frac{k}{\ii\pi} \int_{{-}\infty}^\infty \dd q \, \hat{h}^+_l(qr)
  \frac{\left(q/k\right)^{l}}{q^2-k^2} \\
 &= \frac{k}{\ii\pi}\,(2\pi\ii)\,\hat{h}^+_l(kr)\frac{\left(k/k\right)^{l}}{k+k}
 = \hat{h}^+_l(kr) \,,
\end{spliteq}
\end{widetext}
where we have used the relation~\cite[Eqs.~10.47.10,15]{DLMF}
\begin{spliteq}
 \hat{j}_l(qr) &= \frac{1}{2\ii}\left[\hat{h}^+_l(qr)-\hat{h}^-_l(qr)\right] \\
 &= \frac{1}{2\ii}\left[\hat{h}^+_l(qr)-(-1)^l \hat{h}^+_l(-qr)\right]
\end{spliteq}
in the second step and used the residue theorem to evaluate the final integral.\footnote{%
Note that the residue theorem is applied for the $\Ip(k)>0$ case.
This is because we only construct RH functions that are square-integrable, and
$\hat{h}^+_l(kr)$ is square-integrable if and only if $k$ is in the upper half-plane.}

\begin{figure}[htb]
    \centering
    \includegraphics[width=0.49\textwidth]{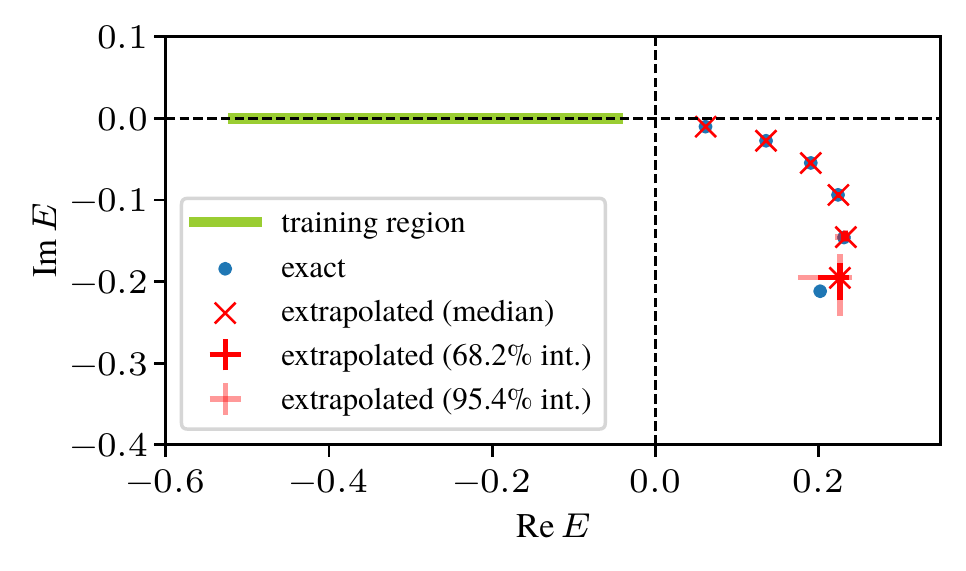}
    \caption{Bound-state-to-resonance extrapolation for the system given by Eq.~\eqref{eq:V_for_s_wave}, performed by augmenting the basis with RH functions with
    wave numbers the same as those of CA-EC vectors.
    Five training points were randomly drawn from the region $c\in (0.9,1.3)$ per dataset.
    See text for details.}
    \label{fig:RHAEC}
\end{figure}

The same potential as given in Eq.~\eqref{eq:V_for_s_wave} in the $S$ wave is used to
test augmenting the basis with RH functions.
As shown in Fig.~\ref{fig:RHAEC}, RH augmentation, like CA-EC, is able to provide bound-state-to-resonance extrapolations, in contradistinction to the naive EC approach.
Note that we use RH augmentation here only to explain why CA-EC
works as well as it does.
Beyond the simple two-body systems we consider here as proofs of concept, RH augmentation would be difficult to implement due to the more complicated structure of few- and many-body wave functions.
CA-EC, on the other hand, is straightforward to implement even for such systems.

%%%%%%%%%%%%%%%%%%%%%%%%%%%%%%%%%%%%%%%%%%%%%%%%%%%%%%%%%%%%%%%%%%%%%%%%%%%%%%%%%%%%%%%%%
%%%%%%%%%%%%%%%%%%%%%%%%%%%%%%%%%%%%%%%%%%%%%%%%%%%%%%%%%%%%%%%%%%%%%%%%%%%%%%%%%%%%%%%%%

\paragraph*{Convergence of basis augmentation with RH functions.}

\begin{figure}[htb]
    \centering
    \RHAECsubfig{(a) using one RH vector}{\includegraphics[width=0.49\textwidth]{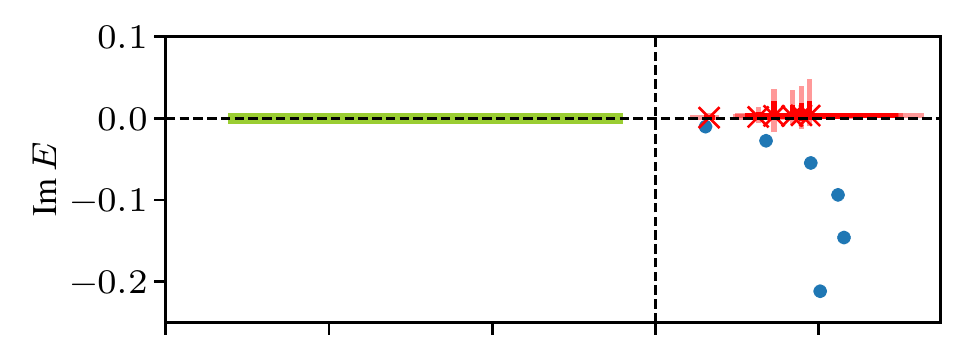}} \\
    \RHAECsubfig{(b) using two RH vectors}{\includegraphics[width=0.49\textwidth]{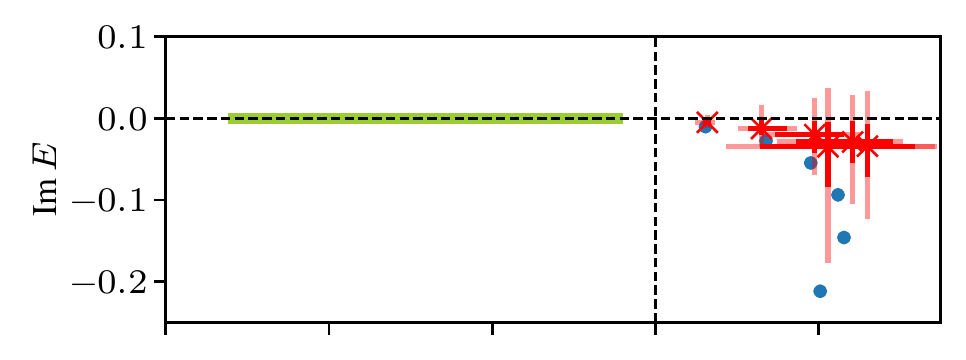}} \\
    \RHAECsubfig{(c) using three RH vectors}{\includegraphics[width=0.49\textwidth]{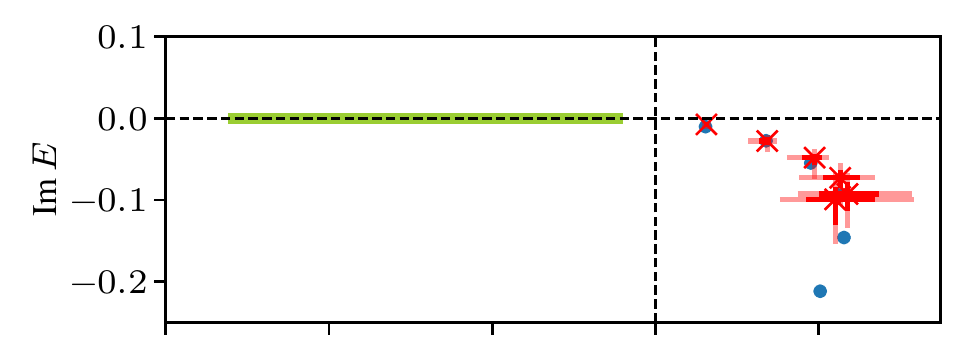}} \\
    \RHAECsubfig{(d) using four RH vectors}{\includegraphics[width=0.49\textwidth]{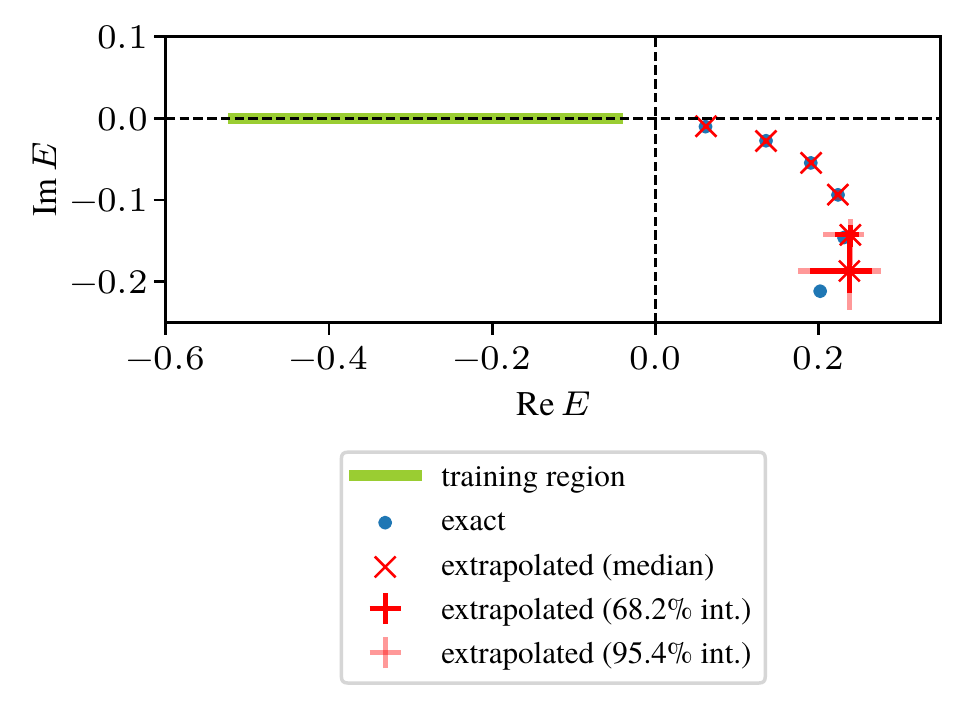}}
    \caption{Bound-state-to-resonance extrapolation for the system given by Eq.~\eqref{eq:V_for_s_wave},
    performed using bases constructed
    with five training points and increasing number of RH vectors.
    The training points were randomly drawn from the region $c\in (0.9,1.3)$.
    From top to bottom, bases are augmented with one, two, three, and four RH vectors, respectively.
    See text for details.}
    \label{fig:RHAEC_convergence}
\end{figure}

To quantify the contribution of RH functions to bound-state-to-resonance extrapolations, we show in Fig.~\ref{fig:RHAEC_convergence} the convergence of extrapolated energies as a function of the number of RH functions added.
For each subplot (in descending vertical order), we have added one, two, three, and four RH functions picked randomly following the same prescription as in the previous calculation.
This calculation was performed for the same system as considered before, given by Eq.~\eqref{eq:V_for_s_wave}.
The rapid convergence with the number of vectors clearly supports the argument that CA-EC provides the asymptotic parts necessary to describe the long-distance structure of resonance wave functions.

%%%%%%%%%%%%%%%%%%%%%%%%%%%%%%%%%%%%%%%%%%%%%%%%%%%%%%%%%%%%%%%%%%%%%%%%%%%%%%%%%%%%%%%%%
%%%%%%%%%%%%%%%%%%%%%%%%%%%%%%%%%%%%%%%%%%%%%%%%%%%%%%%%%%%%%%%%%%%%%%%%%%%%%%%%%%%%%%%%%

\section{Discussion and outlook}
\label{sec:Conclusion}

In this work, we have studied the application of eigenvector continuation to decaying resonance states.
Specifically, we considered a two-body system with a Hamiltonian controlled by a single parameter, which can be tuned to map out resonance trajectories in the complex plane.
Using the uniform complex-scaling technique we showed that
eigenvector continuation can be set up with resonance states as training data to produce an
emulator that predicts resonance properties outside the training domain, provided that the appropriate
c-product is used to construct the EC Hamiltonian and norm matrices.

We demonstrated that a naive implementation of EC trained with only bound states cannot reliably predict resonance properties, even if the bound-state wave functions are defined along
a contour rotated into the complex plane.
We identified this failure to be caused by the lack of outgoing asymptotic behavior in the naive EC basis.
However, we subsequently showed that this problem can be overcome by adding to the EC basis the complex conjugates of the rotated bound-state wave functions, producing an approach that we call conjugate-augmented eigenvector continuation (CA-EC).
Adding the complex-conjugated wave functions provides basis vectors that
effectively give contributions in the \nth{1} quadrant of the complex-momentum plane, which is where the decaying resonances ``exposed'' by the complex-scaling procedure are situated.

We showed with numerical examples that CA-EC provides accurate predictions for decaying resonances at a relatively moderate increase in computational cost compared to standard eigenvector continuation.
We also confirmed the mechanism at play behind the success of the method by replacing the complex-conjugated eigenvectors with Riccati-Hankel functions that provide the same outgoing asymptotic behavior that in CA-EC
is provided by the complex-conjugated bound-state wave functions.

The findings presented in this work provide an important step towards understanding robust bound-state-to-resonance extrapolations, a new tool for the study of open quantum systems in the context of few-body physics, and they open interesting avenues for more efficient many-body applications in the complex-energy plane.
In future work, we will investigate how to leverage the CA-EC method in many-body techniques based on the Berggren basis to emulate multiparticle resonances, with particular applications to exotic atomic nuclei.
We will also consider using more sophisticated methods to optimize the number and the quality of training eigenvectors.

\begin{acknowledgments}
We thank Andrew Andis and Hang Yu for useful discussions and valuable comments about the project.
This work was supported in part by the National Science Foundation under Grant
No.~PHY--2044632.
This material is based upon work supported by the U.S. Department of Energy,
Office of Science, Office of Nuclear Physics, under the FRIB Theory Alliance,
Award No.~DE-SC0013617.
\end{acknowledgments}

\bibliographystyle{apsrev4-1}

\end{document}